\documentclass[a4paper]{article}

\usepackage{INTERSPEECH2022}

\title{Conditional variational autoencoder to improve neural audio synthesis for polyphonic music sound}
\name{Seokjin Lee$^{1,2}$, Minhan Kim$^2$, Seunghyeon Shin$^2$, Daeho Lee$^2$, Inseon Jang$^3$, and Wootaek Lim$^3$}
\address{
  $^1$School of Electronics Engineering, Kyungpook National University\\
  $^2$School of Electronic and Electrical Engineering, Kyungpook National University\\
  $^3$Electronics and Telecommunications Research Institute}
\email{\{sjlee6, kmh7576\}@knu.ac.kr, sineva123@gmail.com, eogh7458@naver.com, \\\{jinsn, wtlim\}@etri.re.kr}

\begin{document}

\maketitle
\begin{abstract}
Deep generative models for audio synthesis have recently been significantly improved. However, the task of modeling raw-waveforms remains a difficult problem, especially for audio waveforms and music signals. Recently, the realtime audio variational autoencoder (RAVE) method was developed for high-quality audio waveform synthesis. The RAVE method is based on the variational autoencoder and utilizes the two-stage training strategy. Unfortunately, the RAVE model is limited in reproducing wide-pitch polyphonic music sound. Therefore, to enhance the reconstruction performance, we adopt the pitch activation data as an auxiliary information to the RAVE model. To handle the auxiliary information, we propose an enhanced RAVE model with a conditional variational autoencoder structure and an additional fully-connected layer. To evaluate the proposed structure, we conducted a listening experiment based on multiple stimulus tests with hidden references and an anchor (MUSHRA) with the MAESTRO. The obtained results indicate that the proposed model exhibits a more significant performance and stability improvement than the conventional RAVE model.
\end{abstract}
\noindent\textbf{Index Terms}: audio synthesis, variational autoencoder, adversarial network.

\section{Introduction}

In recent audio signal processing systems, deep learning-based analysis and synthesis methods exploit various applications related to speech generation, musical composition, and audio compression. Regarding the analysis/synthesis models with deep neural network structures, algorithms have been developed for audio signal generation \cite{chung2015recurrent,dhariwal2020jukebox} and timbre transfer \cite{mor2018universal}. Unfortunately, previously developed models require considerably large dimensionality to model audio waveforms precisely, and are limited in generating high-quality audio waveforms. The recently developed model, differentiable digital signal processing (DDSP) \cite{engel2020ddsp}, exhibits advanced reconstruction performances in terms of quality and naturalness of generated audio waveforms; however, it utilizes a number of pre-defined audio descriptors, which restricts the types of signals that can be generated.    

In the speech signal processing, analysis/synthesis methods have been developed based on simple autoencoder (AE) \cite{kankanahalli2018end} or cascading AE structures \cite{zhen2019cascaded}. In addition, generative model-based speech techniques, such as wavenet and variational autoencoder (VAE) have been successfully developed \cite{kleijn2018wavenet}, and VAE-based techniques are further developed to vector quantization-based VAE (VQ-VAE) \cite{garbacea2019low}. Although these methods targeting a speech signal can optimally reconstruct the waveform with a latent vector of a relatively small complexity, it is severely difficult to apply these methods to audio or music applications. Recently, although the residual VQ (RVQ) model that employs multiple VQ units has attempted to extend the speech model to the music application, it did not obtain sufficient reconstruction quality. 

Recently, the realtime audio variational autoencoder (RAVE) \cite{caillon2021rave} was developed for high-quality audio synthesis. To address the limitations in the existing researches and obtain optimal synthetic quality, the RAVE method introduces the two-stage training procedure, which are representation learning and adversarial fine-tuning stages. In the representation learning stage, the RAVE trains the encoder and decoder networks in VAE modules, and then the decoder is fine-tuned in the adversarial generation stage. Although the RAVE method can generate audio waveforms with a relatively good quality, it is faced with challenges, in that the generated sound quality in the low-frequency band is degraded relative to that in the mid-frequency band, and sometimes, unwanted pitched noise is generated. The former creates problems for listeners, such as missing low-pitched sounds or raising the pitch by one octave (we call this the "missing bass" problem). Furthermore, the latter also causes the listener to hear strange pitched notes, which are more disturbing than typical unpitched noise.

In this study, we proposed an enhanced RAVE structure to solve the missing bass and the pitched noise problem by modifying neural network structures and utilizing pitch information, without any additional assumption (such as weighted loss function for low frequency components). The proposed structure utilizes the conditional VAE (CVAE) scheme with pitch information, such that the model can focus on frequency bands suitable for target pitch when generating audio waveforms.

\begin{figure*}[t]
  \centering
  \includegraphics[width=\linewidth]{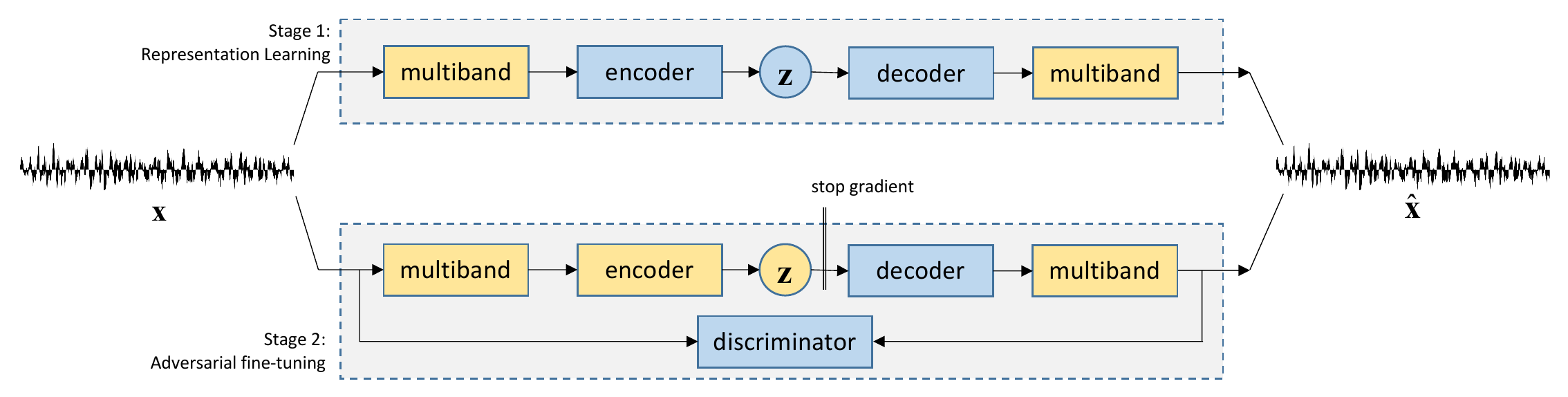}
  \caption{Network structure of the RAVE model. Blue-colored blocks are optimized during each training step, while yellow-colored blocks are fixed or frozen. For details on the structure, refer to \cite{caillon2021rave}}
  \label{fig:RAVE}
\end{figure*}

\section{Preliminary: Realtime Audio Variational Autoencoder}

The development of the RAVE method was inspired by recent achievements in VAE and autoencoding raw waveform models. The VAE is a generative model that attempts to understand data $\mathbf{x} \in \mathbb{R}^{d_{x}}$ by modeling the distribution $p\left( \mathbf{x} \right)$ with latent variables $\mathbf{z} \in \mathbb{R}^{d_{z}}$ \cite{caillon2021rave}. To achieve this objective, the VAE introduces an inference model $q_{\phi} \left( \mathbf{z} \vert \mathbf{x} \right)$ that is optimized to minimize the evidence lower bound (ELBO) as \cite{kingma2014auto}
\begin{equation}
\mathcal{L}_{\phi,\theta} = - \mathbb{E}_{q_{\phi} \left( \mathbf{z} \vert \mathbf{x} \right)} \left[ \log p_{\theta} \left( \mathbf{x} \vert \mathbf{z} \right) \right] + \mathcal{D}_{\mathrm{KL}} \left[ q_{\phi} \left( \mathbf{z} \vert \mathbf{x} \right) \Vert p \left( \mathbf{z} \right) \right],
\label{eq:ELBO}
\end{equation}
where $\mathcal{D}_{\mathrm{KL}}\left( \mathbf{a} \Vert \mathbf{b} \right)$ represents the Kullback-Leibler divergence between $\mathbf{a}$ and $\mathbf{b}$. The posterior distributions $q_{\phi}$ and $p_{\theta}$ are calculated by the encoder and decoder, respectively, with $\phi$ and $\theta$ parameters. Optimizing ELBO minimizes the reconstruction error (the first term of (\ref{eq:ELBO})) with latent regularization (the second term of (\ref{eq:ELBO})).

Because the objective of the RAVE method is to reconstruct the raw waveforms, it is necessary to set a cost function suitable for this purpose. In previous studies on synthesizing audio waveform with feed-forward convolutional autoencoder networks, a perceptually motivated distance measure called \emph{spectral distance} was proposed as \cite{defossez2018sing}
\begin{align}
	\mathcal{D}_{spec} \left( \mathbf{x}, \mathbf{y} \right) = & \lVert \log \left( \lvert \mathrm{STFT} \left( \mathbf{x} \right) \rvert ^{2} + \epsilon \right) \nonumber \\
	& \qquad - \log \left( \lvert \mathrm{STFT} \left( \mathbf{y} \right) \rvert ^{2} + \epsilon \right)\rVert _{1},
\label{eq:spec}
\end{align}
where $\mathrm{STFT} (\mathbf{x})$ means a short-time Fourier transform of $\mathbf{x}$, $\lvert \cdot \rvert$ means an element-wise absolute operator, $\lVert \cdot \rVert_{1}$ means the $L_{1}$-norm, and $\epsilon$ is an arbitrary small value.

The RAVE structure is developed based on the VAE structure. As aforementioned, the first term of ELBO (\ref{eq:ELBO}) is related to the reconstruction error. Considering the previous studies related to the generation of audio waveform, RAVE minimizes a modified ELBO in which the first term is replaced by the spectral distance as
\begin{equation}
\mathcal{L}_{\mathrm{vae}} = - \mathbb{E}_{\mathbf{\hat{x}} \sim p\left( \mathbf{x} \vert \mathbf{z} \right)} \left[ \mathcal{D}_{ms} \left( \mathbf{x}, \mathbf{\hat{x}} \right) \right] + \beta \mathcal{D}_{\mathrm{KL}} \left[ q_{\phi} \left( \mathbf{z} \vert \mathbf{x} \right) \Vert p \left( \mathbf{z} \right) \right],
\label{eq:modELBO}
\end{equation}
where $\mathbf{\hat{x}}$ denotes the generated waveform by RAVE. The multiscale spectral distance \cite{engel2020ddsp}, which is defined as
\begin{align}
\mathcal{D}_{ms} \left( \mathbf{x}, \mathbf{y} \right) = & \sum_{n \in \mathcal{N}} \bigg[ \frac{ \lVert \mathrm{STFT}_{n} \left( \mathbf{x} \right) - \mathrm{STFT}_{n} \left( \mathbf{y} \right) \rVert _{F} }{ \lVert \mathrm{STFT}_{n} \left( \mathbf{x} \right) \rVert _{F}} \nonumber\\ 
& \phantom{1} + \log \left( \lVert \mathrm{STFT}_{n} \left(  \mathbf{x} \right) - \mathrm{STFT}_{n} \left( \mathbf{y} \right) \rVert _{1} \right) \bigg]
\label{eq:multispec}
\end{align}
is utilized as the reconstruction error measure, instead of the simple spectral distance (\ref{eq:spec}). $\mathrm{STFT}_{n} \left( \mathbf{x} \right)$ represents the magnitude of the short-time Fourier transform with window length of $n$ and the hop size of $n/4$, $\mathcal{N}$ is a set of window length parameters, while $\lVert \cdot \rVert _{F}$ denotes the Frobenius norm.


In the audio waveform generation task, two perceptually similar audio signals can contain subtle phase differences, and their differences can significantly alter the waveforms. To address this problem, the training procedure of RAVE comprises two stages: the \emph{representation learning} and \emph{adversarial fine-tuning}, as illustrated in Figure~\ref{fig:RAVE}. In the \emph{representation learning} stage, both the encoder and decoder of the VAE are trained by optimizing the ELBO (\ref{eq:modELBO}). In the \emph{adversarial fine-tuning} stage, the decoder is fine-tuned by adopting the \emph{discriminator} $D$, while the encoder is fixed. To train the decoder and discriminator, a generative adversarial network (GAN) objective is defined as \cite{caillon2021rave}
\begin{align}
\mathcal{L}_{\mathrm{dis}} \left( \mathbf{x}, \mathbf{z} \right) = & \left[ 1 - D(\mathbf{x}) \right]_{+} + \mathbb{E}_{\mathbf{\hat{x}} \sim p\left( \mathbf{x} \vert \mathbf{z} \right)} \left[ 1 + D(\mathbf{\hat{x}}) \right] _{+} ,\\
\mathcal{L}_{\mathrm{dec}} \left( \mathbf{x}, \mathbf{z} \right) = & \nonumber \\ 
\mathbb{E}_{\mathbf{\hat{x}} \sim p\left( \mathbf{x} \vert \mathbf{z} \right)} & \left[ -D(\mathbf{\hat{x}}) + \mathcal{D}_{ms} \left( \mathbf{x}, \mathbf{\hat{x}} \right) + \mathcal{L}_{\mathrm{FM}} \left( \mathbf{x}, \mathbf{\hat{x}} \right) \right],
\end{align}
where $\left[ \cdot \right]_{+}$ denotes a truncation operator of the negative value to zero, and $\mathcal{L}_{\mathrm{FM}}$ represents the feature matching distance proposed by Kumar et al. \cite{caillon2021rave,kumar2019melgan}.

\begin{figure}[t]
	\centering
	\includegraphics[width=\linewidth]{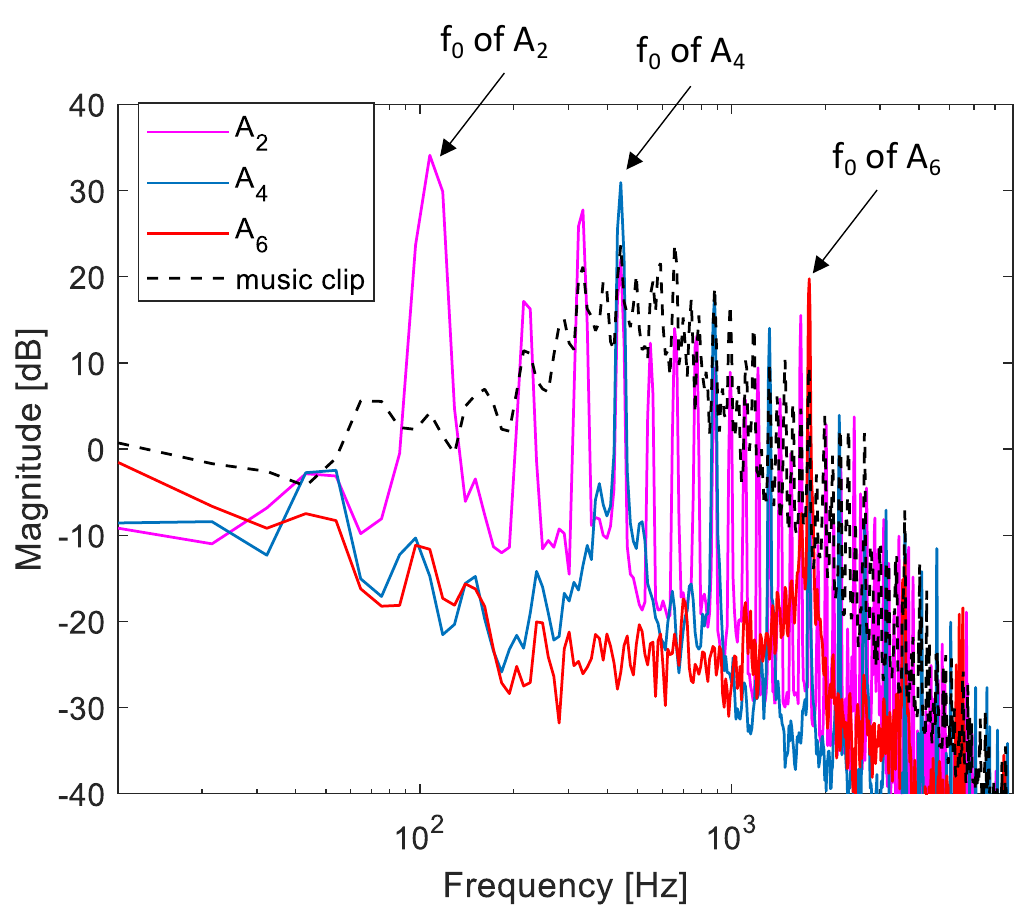}
	\caption{Spectrum of A$_2$(magenta), A$_4$(blue), A$_6$(red) piano notes, and an entire music clip with piano (black dashed line). Each note sound is extracted from \emph{instrument sound} of the RWC music database \cite{goto2004development}, while the piano music clip is obtained from the MAESTRO database \cite{hawthorne2018enabling}.  }
	\label{fig:spectrum}
\end{figure}

\begin{figure*}[t]
	\centering
	\includegraphics[width=14cm]{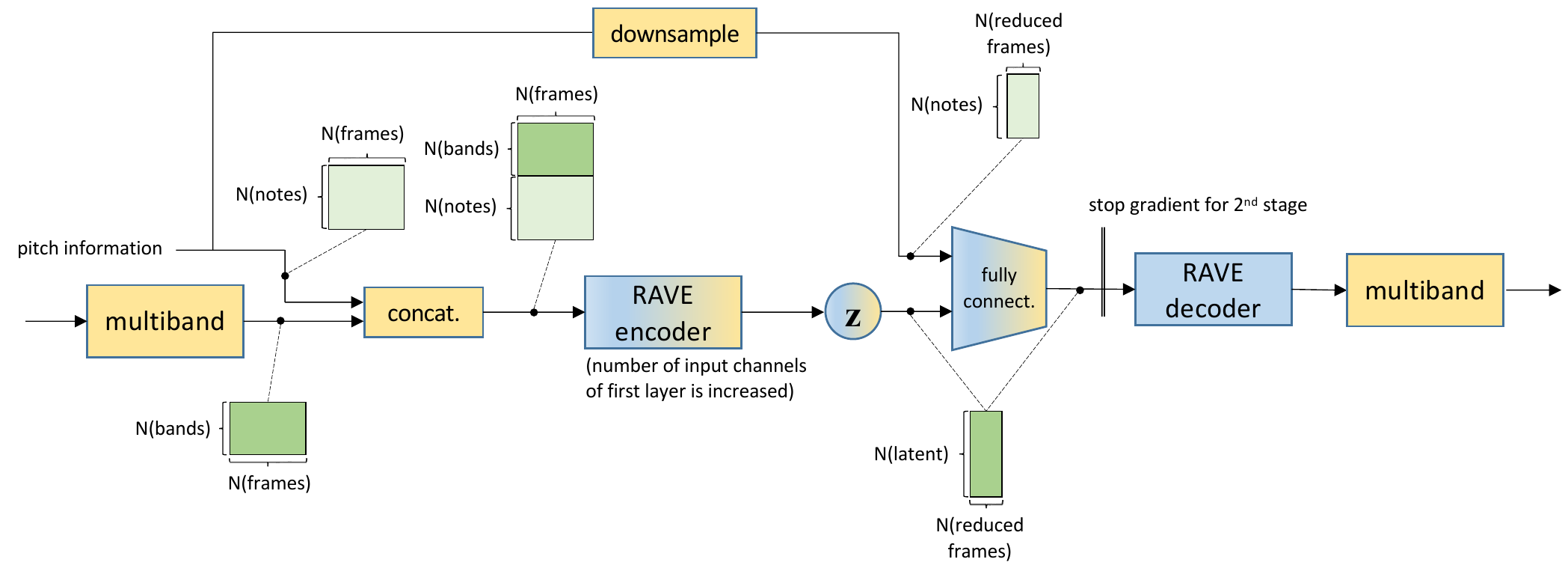}
	\caption{Schematic diagram for proposed modification. Green-colored boxes depict the shapes of the data, and the graduation color from blue to yellow implies the blocks that are optimized during the first stage and frozen during the second stage.}
	\label{fig:mod}
\end{figure*}

\section{Proposed Structure}

\subsection{Motivation}

The RAVE method generates audio signals well in the midrange frequency band (e.g. right-hand side of the piano), and the generated results are natural to the ears. However, the generation performance of RAVE deteriorates in the low and high frequency bands. In particular, in the case of low-pitched sound (e.g. left-hand side of the piano), the low-order harmonics and fundamental frequency component are not optimally reproduced; hence, the low-pitched notes sounds are omitted.

To solve this \emph{missing bass} problem, we consider the spectral distribution of the input/target music signal. As depicted by the solid line in Figure~\ref{fig:spectrum}, the frequency spectrum of a note is concentrated on the fundamental frequency and low-order harmonics, while the entire music signal exhibits a broadband distribution, as depicted by the dotted line in Figure~\ref{fig:spectrum}. In particular, many of the music signals we analyzed exhibited a distribution concentrated in the midrange frequency band, as presented in Figure~\ref{fig:spectrum}. Although it is not yet clearly analyzed, it appears that the range of fundamental frequency of the well-reconstructed notes is related to the frequency band, with the spectral distribution of the input/target music signal. 

Because it is significantly difficult to completely reconstruct all frequency bands at once, we modify the RAVE model to focus on reconstructing the important frequency range, instead of trying to reconstruct all frequency bands. The important frequency range depends on the pitch of each note, as illustrated in Figure~\ref{fig:spectrum}. Therefore, we apply the CVAE structure \cite{sohn2015learning} to the RAVE model. The original CVAE in \cite{sohn2015learning} is conditioned by output variables; however, in subsequent studies \cite{montserrat2019class}, it is also modified to be conditioned by auxiliary information input, similar to the conditional generative adversarial network (CGAN) \cite{mirza2014conditional}. Inspired by the latter research, we propose a CVAE structure for applying the musical instrument digital interface (MIDI) note number as the auxiliary inputs for the encoder and decoder.

\subsection{Proposed modification for the RAVE}

Our modification for the RAVE structure is presented in Figure~\ref{fig:mod}. In the conventional RAVE method, each block of the audio input signal is decomposed by pseudo-quadrature mirror filters into multiband signals such as the $N_{bands} \times N_{frames}$ matrix, which is provided as the input of the encoder. In our modification, the auxiliary encoder data, which is a $N_{notes} \times N_{frames}$ matrix, comprises the one-hot encoded note number of each frame. $N_{notes}$ is set to the possible number of notes, e.g., 88 for the piano music. Because the encoder network comprises a number of convolutional neural networks with major strides, the temporal dimension of the encoder output (i.e. the temporal dimension of the decoder input) is smaller than that of the encoder input. Hence, the auxiliary decoder data is merged to fit the reduced temporal dimension.

Referring to previous studies for conditional generative models such as CGAN \cite{mirza2014conditional}, the auxiliary encoder data is concatenated with the multiband audio data, as illustrated in Figure~\ref{fig:mod}. On the decoder side, the auxiliary data is concatenated with the latent vector and fed into the fully connected layer before being fed to the decoder. The structures of the encoder and decoder are the same as that of the conventional RAVE; however, the number of input channels of the first layer of the encoder is increased by $N_{notes}$. In our experiment, applying the fully connected layer performs better than simply concatenating the auxiliary decoder data to the latent vector, which will be more comprehensively discussed in the next chapter. The fully connected layer for the auxiliary decoder data is trained during the \emph{representation learning} stage and fixed during the \emph{adversarial fine-tuning} stage. The multiband, RAVE encoder/decoder, and discriminator (which is not shown in Figure~\ref{fig:mod}) are same as the conventional RAVE; however, the number of input channels of the first layer of the RAVE encoder is increased by the number of notes. The detailed structures of the encoder/decoder modules can be found in \cite{caillon2021rave}.

\section{Experiment}

\subsection{Experiment and training settings}

\textbf{Datasets.} To evaluate the proposed structure, we trained the enhanced RAVE model using the 2014, 2015, and 2017 data of the MAESTRO dataset \cite{hawthorne2018enabling}. The training data comprised 374 piano clips (approximately 55 h in total), of which 5 \% of the randomly selected data was adopted as the validation data. All audio clips were downsampled to 16 kHz, and down-mixed to mono signal.

\textbf{Training.} The proposed model was trained with 1M and 2M iterations for the \emph{representation learning} and \emph{adversarial fine-tuning} stages, respectively. We adopted the Adam optimizer \cite{kingma2014adam} with a learning rate of $10^{-4}$ where $\beta_1$ and $\beta_2$ were 0.5 and 0.9, respectively. The batch size was set to 8, and the input data was augmented by random crop, all-pass filters with random phase, and dequantization \cite{caillon2021rave}. The auxiliary inputs for encoders and decoders for each frame were extracted as one-hot encoded vectors with 88 elements from the MIDI data included in the dataset. 

\textbf{Evaluation.} The proposed model was evaluated with randomly extracted 10 samples from the 2013 data of the MAESTRO dataset, which included unseen data for training. We compared the reconstructed waveform of the proposed model with that of the conventional RAVE model via multiple stimulus tests with hidden references and anchor (MUSHRA) listening tests. During the test, we asked nine listeners to assess the stimuli similar to the reference among the proposed model, a simple application of CVAE to the RAVE (without the fully connected layer for the auxiliary decoder input), and the conventional RAVE model, in terms of the recognized pitch, fidelity, and naturalness of the generated sound. We adopted the 1-kHz high-pass filtered signal as the anchor, considering the \emph{missing bass} problem of the conventional RAVE model.

\subsection{Experiment results}

\begin{figure}[t]
	\centering
	\includegraphics[width=\linewidth]{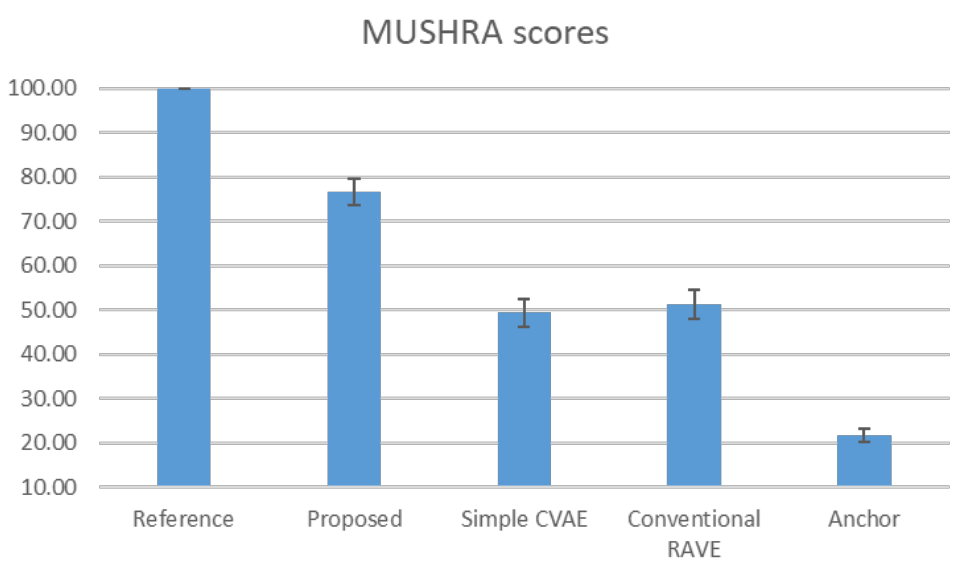}
	\caption{Results of the listening test.}
	\label{fig:MUSHRA}
\end{figure}

Figure~\ref{fig:MUSHRA} presents the MUSHRA scores of each method. The rectangle block and vertical bar denote the averaged score and 95 \% confidence interval of each test model in Figure~\ref{fig:MUSHRA}, respectively. The simple CVAE model, which is the same structure as the proposed model without the fully connected layer for the auxiliary decoder input, does not exhibit any performance improvement compared to the conventional RAVE model, despite the adoption of pitch information as an auxiliary. The averaged scores of the simple CVAE and conventional RAVE are 49.4 and 51.4, respectively, and most of the confidence intervals overlap. However, the averaged score of the proposed model is 76.7, which indicates a significant improvement, considering the confidence interval.

\begin{figure}[t]
	\centering
	\includegraphics[width=\linewidth]{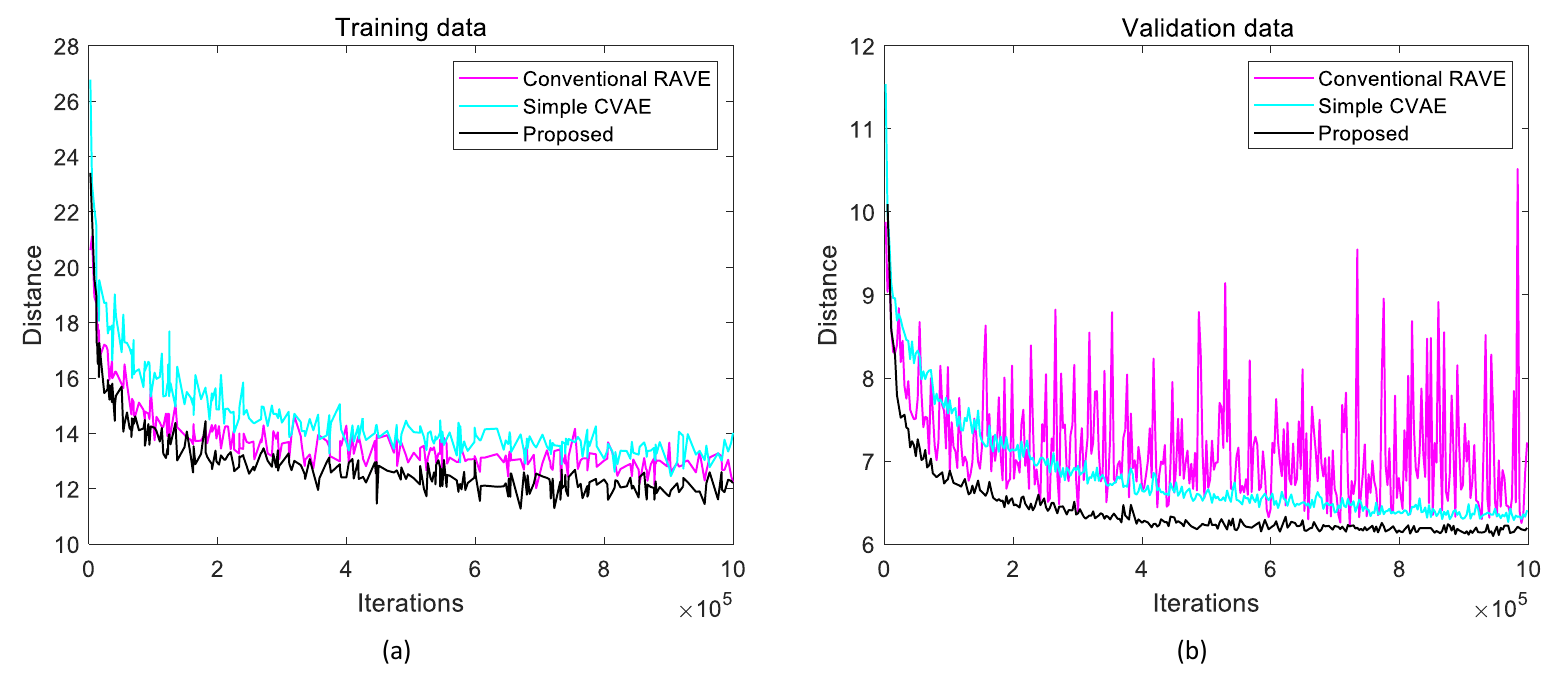}
	\caption{Change in the distance measure between the input and reconstructed waveforms for the (a) training and (b) validation sets during the \emph{representation learning} stage.}
	\label{fig:learning}
\end{figure}

Figure~\ref{fig:learning} (a) and (b) demonstrate that the distances between input and reconstructed waveforms for the training and validation sets, respectively, during the \emph{representation learning} stage. For the training data, the performances of the proposed, simple application of CVAE, and conventional RAVE models are similar. The proposed model exhibits slightly better performance than the conventional RAVE model; however, the difference is not significant. However, the proposed model exhibits a significantly improved and stable performance compared to the conventional model for the validation data.

\begin{figure}[t]
	\centering
	\includegraphics[width=\linewidth]{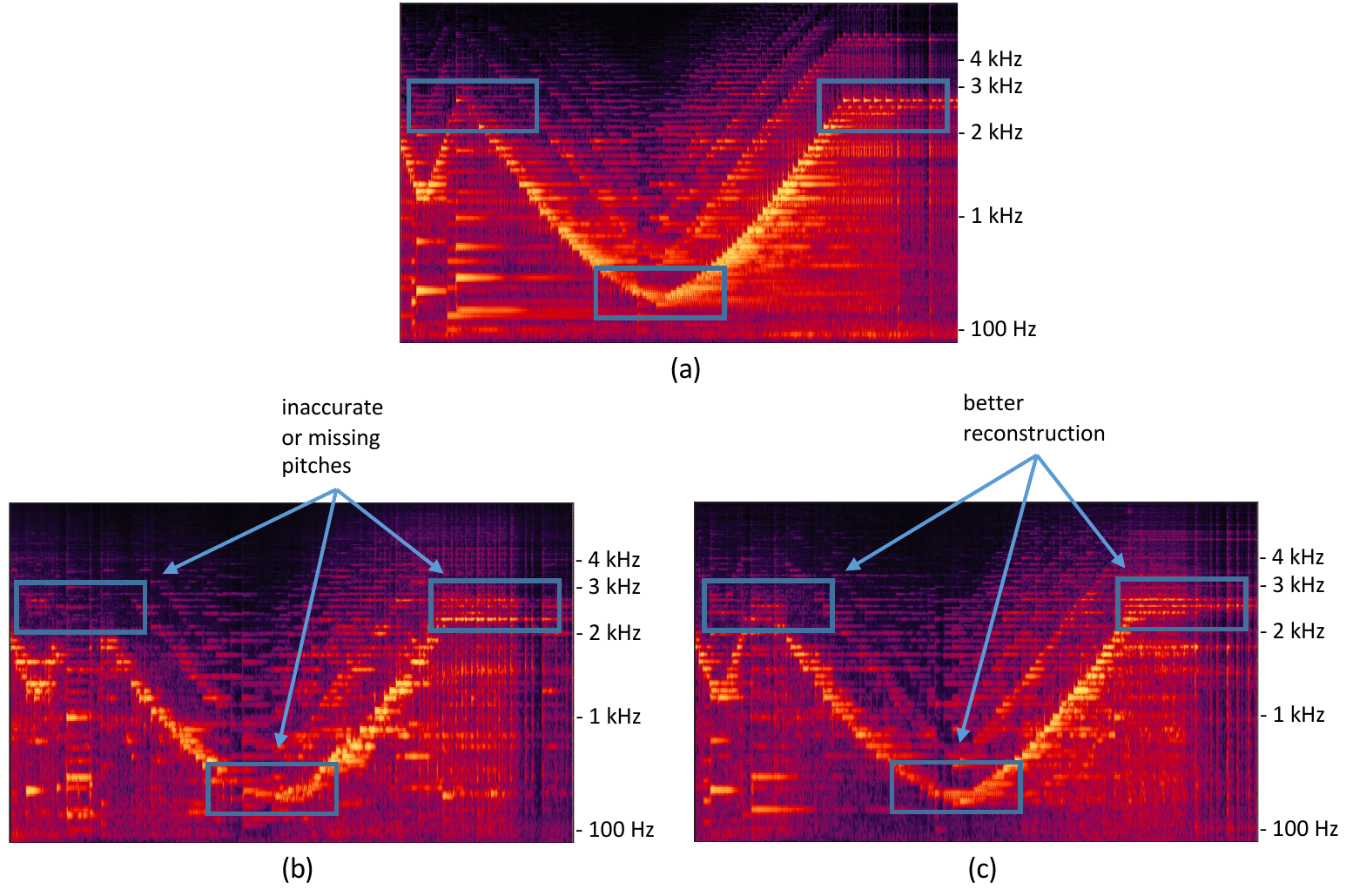}
	\caption{Examples of the spectrograms for (a) the target waveform, reconstructed waveforms of (b) the conventional RAVE, and (c) the proposed model.}
	\label{fig:spectrogram}
\end{figure}

When the listener hears the reconstructed waveforms of the proposed and the conventional RAVE models, the difference is more significant than the difference in the distance measures. This is because the \emph{missing bass} problem in the conventional model has been addressed, thus allowing listeners to recognize the pitches clearly. Figure~\ref{fig:spectrogram} presents spectrograms of the reference and reconstructed signals of the conventional RAVE and proposed model. In this audio clip, the pitch decreases and increases rapidly, thus forming a V-shaped spectrogram, as illustrated in Figure~\ref{fig:spectrogram}(a). Comparing the results of the conventional and the proposed models, the performance improvement is particularly substantial in low- and high-pitched notes (indicated by blue boxes).

\section{Conclusion}

In this study, an enhanced model for audio synthesis was developed based on the RAVE. The proposed model utilizes pitch information as an auxiliary information, to solve the \emph{missing bass} problem and help the network to focus on the reconstruction of the important band. To utilize the pitch information, the CVAE structure is applied to the RAVE-based model, and an additional fully-connected layer is added to handle the auxiliary decoder input. The subjective listening test based on the MUSHRA method indicates that the proposed model exhibits a more significantly enhanced performance than the conventional RAVE. In addition, the proposed model is advantageous in reconstructing the fundamental frequency and low-order harmonics of low- and high-pitched sound. Certainly, the proposed model requires pitch information; however, automatic music transcription techniques, such as onsets and frames \cite{hawthorne2017onsets}, may help in acquiring the pitch information.

\section{Acknowledgements}
This work was supported by Electronics and Telecommunications Research Institute(ETRI) grant funded by the Korean government.
[22ZH1200, The research of the basic media$\cdot$contents technologies]

\bibliographystyle{IEEEtran}

\bibliography{MusRAVE}

\end{document}